\documentclass[aps,preprint,prd,eqsecnum,nofootinbib,showpacs,12pt]{revtex4}

\usepackage{epsfig}
\usepackage{bm}

\newcommand{\bq}{\begin{equation}}
\newcommand{\be}{\begin{equation}}
\newcommand{\eq}{\end{equation}}
\newcommand{\ee}{\end{equation}}
\newcommand{\bqa}{\begin{eqnarray}}
\newcommand{\ba}{\begin{eqnarray}}
\newcommand{\eqa}{\end{eqnarray}}
\newcommand{\ea}{\end{eqnarray}}
\newcommand{\psl}{p\hskip-0.21cm\slash}
\newcommand{\lsl}{l\hskip-0.21cm\slash}
\newcommand{\esl}{e\hskip-0.21cm\slash}

\begin{document}
\pagestyle{empty}

\title{
{\vspace{-1.2em} \parbox{\hsize}{\hbox to \hsize
{\hss \normalsize\rm PM/03-07}}} \\
Reliability of a high energy one-loop expansion\\
 of $e^+e^- \to W^+W^-$ 
in the SM and in the MSSM
\footnote{Partially supported by EU contract HPRN-CT-2000-00149}}

\author{Matteo Beccaria}
\email{Matteo.Beccaria@le.infn.it}
\affiliation{INFN, Sezione di Lecce, and
Dipartimento di Fisica dell'Universit\`a di Lecce,
Via Arnesano, ex Collegio Fiorini, I-73100 Lecce, Italy}

\author{Fernand M. Renard}
\email{renard@lpm.univ-montp2.fr}
\affiliation{ Physique Math\'{e}matique et Th\'{e}orique, UMR 5825\\
Universit\'{e} Montpellier II,  F-34095 Montpellier Cedex 5.}

\author{Claudio Verzegnassi}
\email{claudio@ts.infn.it}
\affiliation{Dipartimento di Fisica Teorica, Universit\`a di Trieste, \\
Strada Costiera  14, Miramare (Trieste) \\
INFN, Sezione di Trieste
}

\begin{abstract}
We compare the logarithmic Sudakov expansions of the process 
$e^+e^- \to W^+W^-$ in  the one-loop approximation and in the 
resummed version, to subleading order accuracy, in the SM case and 
in a light SUSY scenario for the MSSM. 
We show that the two expansions are essentially identical below 
1 TeV, but differ drastically at higher (2,3 TeV) center of mass energies. 
Starting from these conclusions, we argue that a complete one-loop 
calculation in the energy region below 1 TeV does not seem to need
extra two-loop corrections, in spite of the relatively large size of
the one-loop effects.
\end{abstract}


\maketitle

\pagestyle{plain}

\section{Introduction}

Historically, $WW$ production from $e^+e^-$ annihilation has been a
basic process for the purposes of testing the SM and several of its
extensions or alternatives \cite{WWrev}. This fact has been known for
decades, and has already led to several constraints from the available
measurements at LEP2\cite{LEP2}. From a practical point of view, the
derivation of the constraints is relatively straightforward since in
many cases it
does not require, given the total absence of any signal of new 
physics, an extremely precise theoretical calculation. 
This situation should change 
at the foreseen future linear colliders 
(LC\cite{LC}, CLIC\cite{CLIC}), where
an accuracy at the level of a relative few (1-2) percent 
is expected and,
hopefully, evidence of some kind of new physics will have already
been given. In this spirit, a rigorous calculation of effects 
beyond the simple tree level is a theoretical must.\par
In the simplest case of the SM, this complete one-loop calculation has
been performed almost a decade ago \cite{BD}. Quite recently,
Hahn\cite{Hahn} has redone the SM calculation (finding complete
agreement with Ref.\cite{BD}) and extended it to the MSSM case. His
analysis has been performed in an energy range between, roughly, 200
GeV and 1 TeV and represents, to our knowledge, the only
\underline{complete} analysis of this process in the MSSM 
which is nowadays
available.\par
A feature that one notices at first sight in various plots of the
different considered cross sections is the fact that the one-loop
effects become almost immediately large, reaching a relative size of
fifteen percent or more, both in the SM and in the MSSM (where they
are systematically larger). On one side, this has the positive
important meaning that the radiative corrections do provide a real
stringent test of the model. On another side, this could introduce a
disturbing warning: if a final accuracy at the one percent level is
aimed, a fifteen percent one-loop effect might require the hard
calculation of higher order (e.g two-loop) effects, that seems, at
least in the MSSM case, beyond the available technical (and human)
possibilities.\par
The aim of this paper is that of arguing that, at least at the 
relative few percent level, a one-loop calculation does not seem 
to need extra corrections below 1 TeV, but cannot give the same
reliability at higher (2,3 TeV) center of mass (c.m.) energies. With this purpose, we
shall divide our discussion into three parts. In Section II we shall
compute the electroweak Sudakov logarithmic expansion (for
this terminology see e.g Ref.\cite{sud})
of the process to subleading logarithmic accuracy, both at the one
loop level and resummed to all orders. For c.m. energies beyond a few
hundred GeV this asymptotic expansion should be certainly valid for
the SM. In the MSSM case, a priori, it can be assumed to be reliable
in the picture of a relatively light SUSY scenario, where all the
relevant \underline{electroweak} sparticle masses are below a
few hundred GeV; our analysis will be limited therefore to this
special configuration. In Section III we shall compare the effects on
the different observables of the process of the two approximate
expansions and show that they are "essentially" ({\em i.e.} at the assumed
few percent accuracy level) identical below 1 TeV, but drastically
different at higher (2,3 TeV) energies. From this evidence 
we shall then try to conclude in Section IV
that a complete one-loop calculation 
does not seem to require extra two-loop corrections in the energy
region below 1 TeV, the proposed range of a future LC.

\section{Logarithmic Sudakov expansions}

In all our analysis we shall assume for the purposes of our
logarithmic Sudakov expansions, that are by definition asymptotic
ones, that the involved c.m. energies are sufficiently higher than all
the involved masses, and therefore we shall consider the behaviour of
the process in the "high energy" limit. In this range, we shall now
provide a brief description of the essential kinematical properties of
the scattering amplitude, both at the Born level and at the considered
higher orders.

\subsection{High energy behaviour at the Born level}

Denoting by $l_1,l_2,p_1,p_2$ the $e^-,e^+,W^-,W^+$ momenta
and $e_1,e_2$ the $W^-,W^+$ polarization vectors,
the invariant Born amplitude due to
$\nu_e,~\gamma,~Z$ exchanges is:
\bq
A^{Born}=A^{\nu_e,Born}+A^{\gamma,Born}+A^{Z,Born}
\eq
\noindent
with
\bqa
A^{\nu_e,Born}&=&-~{e^2\over 2s^2_W t}~\bar v(e^+)(\esl_2)
(\psl_1-\lsl_1)(\esl_1)P_L u(e^-)
\eqa
\bqa
A^{\gamma+Z,Born}&=&{2e^2\over s} \bar v(e^+)~
[(e_2.p_1)(\esl_1)-(e_1.e_2)(\psl_1)-(e_1.p_2)(\esl_2)]\times\nonumber\\
&&[(1-\chi{2s^2_W-1\over2s^2_W})P_L+
(1-\chi)P_R)]u(e^-)
\eqa
\noindent
where $P_{R/L}=(1\pm\gamma^5)/2$, $s=(l_1+l_2)^2=(p_1+p_2)^2,~
t=(l_1-p_1)^2
=M^2_W-{s\over2}(1-\beta\cos\vartheta)$,~
$u=(l_2-p_1)^2=M^2_W-{s\over2}(1+\beta\cos\vartheta),~
\beta=\sqrt{1-{4M^2_W\over s}}$,~$\chi=s/( s-M^2_Z)$.

\underline{Helicity amplitudes}\\
 From these invariant forms one obtains the helicity amplitudes 
denoted as $F(\lambda,\mu_1,\mu_2)$.
In the limit of negligible electron mass, $\lambda$ is the
chirality (L when $\lambda=-1$
and $R$ when $\lambda=1$) related to the $e^{\pm}$ helicities
$\lambda\equiv2\lambda(e^-)=-2\lambda(e^+)=\pm1$. The $W^{\pm}$
helicities are $\mu_1(W^-)=\pm1,0$ and $\mu_2(W^+)=\pm1,0$
and we will denote by $W^{\pm}_T$ and $W^{\pm}_0$
the transverse and longitudinal cases. For details see
for example Ref.\cite{BD}. We now just recall the main results that we
will use to write the contributions at one loop and beyond.\par
It is convenient to consider separately the amplitudes corresponding
to the 3 different types of final states,
$W_TW_T$, $W_TW_0$, $W_0W_0$, 
whose properties are, already at Born level, very
different in the high energy limit $s\gg M^2_W$:\\ 

${\bf (W_TW_0)}$:~~ The 
\underline{$W_TW_0$ Born amplitudes} are mass
suppressed ({\em i.e.} behave like $M_W/\sqrt{s}$) and we shall ignore
them.\\

${\bf (W_TW_T)}$:~~ The only non vanishing 
\underline{$W_TW_T$ Born amplitudes}
are given by the t-channel
neutrino exchange diagram and, because
of the  purely Left-handed $We\nu$ coupling,
they reduce to
\bq
F^{Born}(L,\mu,-\mu)\simeq -({e^2\over 2s^2_W})({s\over2t})
\sin\vartheta (\mu-\cos\vartheta)
\eq
\noindent
At high energy these amplitudes are strongly peaked forward as
\bq
{2t\over s}=\beta\cos\vartheta-1+{2M^2_W\over s}\to
\cos\vartheta-1
\eq
\noindent
For the purposes of our asymptotic expansions, we must assume large
values of $s$ \underline{and} $t$. Therefore we shall introduce an
angular cut in our numerical computations as discussed in Section
III.\\
Note that the other set of $W_TW_T$ amplitudes 
$F^{Born}(\lambda,\mu,\mu)$ 
receives contributions from both $\nu_e$, $\gamma$ and $Z$
exchanges whose sum cancels in the high energy limit
\bqa
F^{Born}(\lambda,\mu,\mu)\simeq O(M^2_W/s)
\eqa

${\bf (W_0W_0)}$:~~ \underline{The $W_0W_0$ Born amplitudes} result 
from the addition of $\nu_e$, $\gamma$ and $Z$ exchange contributions
leading to
\bqa
F^{Born}(L,0,0)\simeq - ~{e^2\over4s^2_Wc^2_W} \sin\vartheta
~~~~~~~~
F^{Born}(R,0,0)\simeq ~{e^2\over2c^2_W} \sin\vartheta
\eqa

The cancellations leading to the
above results at order $M^2_W/s$, usually dubbed as "gauge
cancellations" (resulting from the relation between the $We\nu$
coupling and the self-gauge boson $\gamma WW$, $ZWW$ couplings), 
lead to results that are in agreement with the 
equivalence theorem \cite{eqth} which states the equality
between the $e^+e^-\to W^+_0W^-_0$ amplitudes and the
amplitudes for charged Goldstone
boson production $e^+e^-\to G^+G^-$. This equality is true not only at
Born level but also at higher orders at logarithmic accuracy
\cite{revmel}. The high energy behaviour of $G^+G^-$
can be found in  previous papers  \cite{scalar, HH}.\\

\subsection{Amplitudes at one loop}

At one loop the amplitudes at logarithmic accuracy can be
obtained either by a direct computation of the high energy
limit of the Feynman diagrams like in our previous works
see {\em e.g.} \cite{scalar, HH, susylog} or by using
the splitting function formalism~\cite{Splitting} 
and the Parameter Renormalization (PR)
properties giving the leading logarithms arising from soft and
collinear singularities \cite{real}. The results agree
and give

\bqa
{\bf (W_TW_T)}:~~~~~~F(L,\mu,-\mu)&=&F^{Born}(L,\mu,-\mu)~\{~
1+(b^{in}_L)[n_i\log{s\over M^2_W}-\log^2{s\over M^2_W}]\nonumber\\
&&
+b^{fin,TT}[-\log^2{s\over M^2_W}]+
b^{ang,TT}_{L}[\log{s\over M^2_W}]~\}
\label{TTamp}\eqa

\bqa
{\bf (W_0W_0)}:~~~~~~F(\lambda,0,0)&=&F^{Born}(\lambda,0,0)~\{~
1
+b^{in}_{\lambda}[n_i\log{s\over M^2_W}-\log^2{s\over M^2_W}]\nonumber\\
&&
+b^{fin,00}[n_f\log{s\over M^2_W}-\log^2{s\over M^2_W}]
-b^{Yuk,00}[\log{s\over M^2_W}]+
b^{ang,00}_{\lambda}[\log{s\over M^2_W}]\nonumber\\
&&
-~{\alpha
\over\pi}[({c^2_W\over s^2_W}{\tilde \beta}_0 
+{s^2_W\over c^2_W}{\tilde \beta}_0^\prime )\delta_{\lambda,L}
+{{\tilde \beta}_0^\prime\over c^2_W}\delta_{\lambda,R}]
[\log{s\over M^2_W}]~\}
\label{00amp}\eqa
\noindent
where a common mass scale, chosen to be $M_W$, has been used both in
the genuine Sudakov logarithms and in the linear ones of Renormalization
Group (RG) origin
discussed below. With this
choice, the dependence on the MSSM mass parameters has been fully
shifted into residual next-to subleading terms (either constant or
vanishing with energy) that are beyond the purposes of this paper.

In Eqs.(\ref{TTamp}-\ref{00amp}) one recognizes

\begin{enumerate}
\item[---] the \underline{universal} effects
for the incoming $e^+,e^-$  with the form 
$[n_i\log{s\over M^2_W}-\log^2{s\over M^2_W}]$,
$n_i=3$ in SM and $n_i=2$ in MSSM, as established in
Refs.\cite{scalar,real} and 
the coefficients

\bq
b^{\rm in}_{L}=\frac{\alpha(1+2c^2_W)}{16\pi s^2_Wc^2_W}
~~~~~~~~b^{\rm in}_{R}=\frac{\alpha}{4\pi c^2_W}
\eq

\item[---] the \underline{universal} effects
for the outgoing $W^+_T,W^-_T$  with the form
$[-\log^2{s\over M^2_W}]$,
both in SM and in MSSM, \cite{real} 
and 
the coefficient

\bq
b^{\rm fin,TT}=\frac{\alpha}{2\pi s^2_W}
\eq

\item[---] the \underline{universal} effects
for the outgoing $W^+_0\equiv G^+ ,W^-_0\equiv G^-$  with the form
$[n_f\log{s\over M^2_W}-\log^2{s\over M^2_W}]$,
$n_f=4$ in SM and $n_f=2$ in MSSM for what concerns the gauge part
and the additional Yukawa term for both models, with the
coefficients \cite{scalar, HH, real}

\bq
b^{\rm fin,00}= \frac{\alpha(1+2c^2_W)}{16\pi s^2_Wc^2_W}~~~~~~~~~
b^{Yuk,00}= \frac{3\alpha(m^2_t+m^2_b)}{8\pi s^2_WM^2_W}\label{bfin}\eq

\item[---] \underline{the non universal (angular dependent) contribution}
which only consists in residual terms arising
from the quadratic logarithms
$\log^2t,~\log^2u$ (from which the $\log^2s$ part has been
subtracted and put in the universal contribution)
generated by t-channel triangles and box diagrams
containing $W,Z,\gamma$
gauge boson internal lines, leading for both SM and MSSM cases to \\
\bq
b^{\rm ang,TT}_{L}=-~\frac{\alpha}{2\pi s^2_W}
[\log\frac{t}{u}+(1-{t\over u})\log\frac{-t}{s}]
\eq
\noindent
and
\bq
b^{\rm ang,00}_{L}=-~\frac{\alpha}{4\pi}
[{4c^2_W\over s^2_W}\log\frac{-t}{s}
+{1\over s^2_Wc^2_W}\log\frac{t}{u}]
~~~~~~~b^{\rm ang,00}_{R}=-~\frac{\alpha}{2\pi c^2_W}
[\log\frac{t}{u}]\eq\\
\end{enumerate}

In the ($W_0W_0$) case, the last term is the one loop
\underline{single log arising
from the RG} \underline{contribution}
 (intermediate $\gamma,Z$ self-energy
contributions). It can be directly obtained from the corresponding Born
contribution through the expression

\begin{equation}
 F^{\rm RG}=-{1\over4\pi^2}~\left(g^4{\tilde \beta}_0{dF^{\rm Born}
\over dg^2}+
~g^{'4}{\tilde \beta}_0^\prime{dF^{\rm Born}
\over dg^{'2}}~\right)\  \log{s\over M^2_W}
\label{RGder}\end{equation}
 where  $gs_W = 
g'c_W = e$ and
 \bqa
&&{\tilde \beta}_0= \frac{43}{24}-\frac{N}{3}~~~~ 
(\mbox{SM}),~~~~~~~~~~~~~~~~
\frac{5}{4}-\frac{N}{2}~~~~  (\mbox{MSSM})\nonumber\\
&&
{\tilde \beta}_0^\prime=-\frac{1}{24}-\frac{5N}{9}~~~~  (\mbox{SM}),~~~~~~~~~
-\frac{1}{4}-\frac{5N}{6}~~~~  (\mbox{MSSM})
 \label{eq:bMSSM}
\end{eqnarray}
or by taking the first order expansion of the running expressions
 \bqa
 g^2(s) &=& \frac{g^2(\mu^2)}{1+{\tilde \beta}_0 \frac{g^2
 (\mu^2)}{4\pi^2}
 \log \frac{s}{M^2_W}} \;\;,\;
 {g^\prime}^2 (s) = \frac{{g^\prime}^2 (\mu^2)}{1+{\tilde \beta}^\prime_0
 \frac{{g^\prime}^2 (\mu^2)}{4\pi^2}
 \log \frac{s}{M^2_W}} \label{aRG}
 \eqa

In the ($W_TW_T$) case there is no such RG term. This can be seen
either in the diagrammatic way (there is no $\gamma,Z$
self-energy contribution to $F(L,\mu,-\mu)$), or when using the
splitting function formalism and the PR analysis 
by observing the cancellation
of the single log due to the collinear singularities associated
to the final lines with the one arising from the
PR contribution \cite{real}.\\ 

The SM part of the above results agrees completely
with a previous analysis
\cite{dp1}. We have also checked that these high energy
$W_TW_T$ properties are in agreement with those of the 
Wino components in chargino pair production \cite{NR}, due to 
supersymmetry.\\

\subsection{Resummed Amplitudes}

The general procedure for writing the so-called resummed
amplitude, {\em i.e.} the exponential form containing all orders
at subleading logarithmic accuracy has been described in several 
papers \cite{revmel, resum} and applied to specific cases
\cite{scalar,NR}.
It has also been very recently successfully checked by a non trivial
comparison with a partial two-loop calculation to leading order
\cite{twoloop} and
with the angular dependent subleading contribution for arbitrary
processes\cite{DMP}. So we will not repeat it here, but will
immediately write the result in a rather
transparent form.

For $W_TW_T$ amplitudes the expression is:
\bqa
F(L,\mu,-\mu)&=&F^{Born}(L,\mu,-\mu)~{\rm exp}~\{~
[\bar{b}^{in}_{L}+\bar{b}^{fin,TT}]
[{1\over3}\log^3({s\over M^2_W})]\nonumber\\
&&
+(b^{in}_L)[n_i\log{s\over M^2_W}-\log^2{s\over M^2_W}]
+b^{fin,TT}[-\log^2{s\over M^2_W}]\nonumber\\
&&+
b^{WPR} [\log{s\over M^2_W}]+
b^{ang,TT}_{L}[\log{s\over M^2_W}]~\}\nonumber\\
&&+F^{RG}(L,\mu,-\mu)
\label{expTT}\eqa

and for $W_0W_0\simeq G^+G^-$ amplitudes:
\bqa
F(\lambda,0,0)&=&F^{Born}(\lambda,0,0)~{\rm exp}~\{~
[\bar{b}^{in}_{\lambda}+\bar{b}^{fin,LL}]
[{1\over3}\log^3({s\over M^2_W})]\nonumber\\
&&
+b^{in}_{\lambda}[n_i\log{s\over M^2_W}-\log^2{s\over M^2_W}]
+b^{fin,00}[n_f\log{s\over M^2_W}-\log^2{s\over M^2_W}]
\nonumber\\
&&-b^{Yuk,00}[\log{s\over M^2_W}]+
b^{ang,00}_{\lambda}[\log{s\over M^2_W}]~\}\nonumber\\
&&+F^{RG}(\lambda,0,0)
\label{exp00}\eqa
\noindent
with $n_i=3$ or $2$, $n_f=4$ or $2$, as in the one loop case. 
The new quantities arising from Parameter Renormalization
of high order diagrams, not defined in the one loop expression, are:
\bq
\bar{b}^{fin,~TT}=
{\alpha^2\tilde\beta_0\over2\pi^2 s^4_{\rm w}}
~~~~~~~~b^{WPR}={\alpha\tilde\beta_0\over\pi s^2_{\rm w}}
\eq

\bq
\bar{b}^{in}_L= \bar{b}^{fin,~LL}
=\frac{3\alpha^2\tilde\beta_0}{16\pi^2 s^4_{\rm w}}+
\frac{\alpha^2\tilde\beta'_0}{16\pi^2 c^4_{\rm w}}
~~~~~~~~
\bar{b}^{in}_R=
\frac{\alpha^2\tilde\beta'_0}{4\pi^2 c^4_{\rm w}}
\eq

The additional RG contribution to all orders are obtained 
explicitely using Eq.(\ref{aRG}) in the SM or MSSM cases:

\bq
F^{RG}(L,\mu,-\mu)= -({s\over4t})\sin\vartheta (\mu-\cos\vartheta)
~\{~g^2(s)-[g^2~]_{Born}~\}
\eq

\bqa
F^{RG}(L,0,0)= - ~({\sin\vartheta\over4})~\{~g^2(s)
+g'^2(s)-[g^2+g'^2]_{Born}~\}
\eqa

\bqa
F^{RG}(R,0,0)= ~{\sin\vartheta\over2}~\{~g'^2(s)-[g'^2~]_{Born}~\}
\eqa

One can check that a first order expansion of  
Eq.(\ref{expTT},\ref{exp00})
reproduces the one loop results. In particular for the
$W_TW_T$ amplitudes one observes the cancellation of the single log
associated to the outgoing $W^+,W^-$ when adding the part coming
from the exponential and the RG part.

\subsection{Observables }

In this Section we summarize the definition of the various
observables that will be considered in the analysis.\\

The \underline{unpolarized angular distribution} is given by
\bqa
{d\sigma\over d\cos\vartheta}&=&
{\beta\over128\pi s}\sum_{\lambda,\mu,\mu'}
|F(\lambda,\mu,\mu')|^2
\eqa
The angular distribution for longitudinally polarized W is obtained
by reducing the sum over $\mu$, $\mu'$ to the case $\mu=\mu'=0$.\\

The left-right angular distribution is given by
\bqa
{d\sigma_{LR}\over d\cos\vartheta}&=&
{1\over128\pi s}\sum_{\mu,\mu'}
[|F(L,\mu,\mu')|^2-|F(R,\mu,\mu')|^2]
\eqa

The differential cross sections can be integrated with an 
angular cut-off $|\cos\vartheta|~\le~\cos~\vartheta_{cut}$ in the forward
or backward cones:
\bq
\int_F \equiv \int_0^{\cos\vartheta_{cut}} d\cos\vartheta,\qquad
\int_B \equiv \int_{-\cos\vartheta_{cut}}^0 d\cos\vartheta
\eq
This leads to the cross sections
\bqa
\sigma    &=& \left(\int_F+\int_B\right)\frac{d\sigma}{d\cos\vartheta} \\
\sigma_{LR} &=& \left(\int_F+\int_B\right)\frac{d\sigma_{LR}}{d\cos\vartheta} \\
\sigma_{FB} &=& \left(\int_F-\int_B\right)\frac{d\sigma}{d\cos\vartheta}
\eqa
and the related asymmetries
\bq
A_{FB} = \frac{\sigma_{FB}}{\sigma},\qquad A_{LR} = \frac{\sigma_{LR}}{\sigma}
\eq
The relative effect due to radiative corrections (treated in the one-loop
approximation or by the resummation formula) is defined to be 
\bq
\mbox{cross-sections}: \Delta\sigma/\sigma \equiv \frac{\sigma-\sigma^{Born}}{\sigma^{Born}},\qquad
\mbox{asymmetries}: \Delta A \equiv A-A^{Born}
\eq
In the high-energy limit ($s\to\infty$) the Born values of the above quantities 
are ($\delta\equiv\cos\vartheta_{cut}$)
\bqa
\sigma &=& \frac{\pi\alpha^2}{192\ s\ s_W^4\ c_W^4}\left[
-3(23-48s_W^2+20s_W^4)\delta+(-9+16s_W^2-12s_W^4)\delta^3+48c_W^4\log\frac{1+\delta}{1-\delta}
\right] \nonumber \\
\sigma_{FB} &=& \frac{\pi\alpha^2}{8\ s\ s_W^4}\left[
-\delta^2-2\log(1-\delta^2)
\right] \nonumber \\
\sigma_{LR} &=& \frac{\pi\alpha^2}{192\ s\ s_W^4\ c_W^4}\left[
-3(23-48s_W^2+28s_W^4)\delta+(-9+16s_W^2-4s_W^4)\delta^3+48c_W^4\log\frac{1+\delta}{1-\delta}
\right] \nonumber \\
\sigma^{{\rm long.} W} &=& \frac{\pi\alpha^2(1-4s_W^4)}{96\ s\ s_W^4\ c_W^4} \nonumber
\eqa

After this, we hope not too long, technical presentation we are now
ready to move to the explicit comparison of the relative effects in
the two expansions. This will be done in the forthcoming Section III.

\section{Comparison of the two expansions}

\begin{figure}[tb]
\begin{center}
\leavevmode
\epsfig{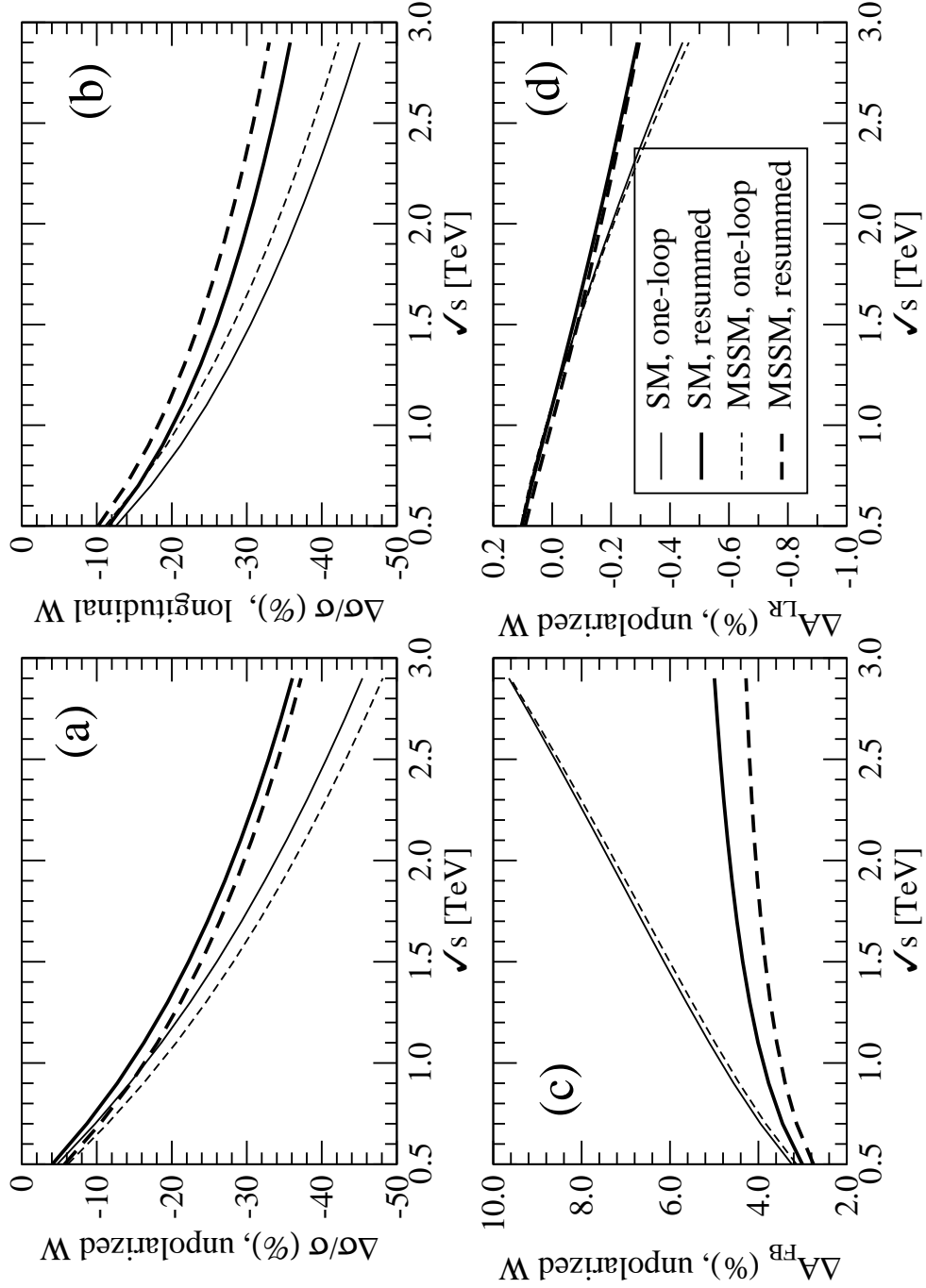}
\vskip-7mm
\end{center}
\caption{Effect of electroweak Sudakov radiative corrections in 
(a) cross-section for production of unpolarized $W$, (b) cross-section for production of 
longitudinally polarized $W$,
(c) forward-backward asymmetry (unpolarized $W$), (d) left-right asymmetry (unpolarized $W$). 
The angular integrations are performed with the cut $\vartheta_{cut}=30^{o}$. For each observable
we show four lines corresponding to the one-loop/resummed 
expansions in the subleading logarithmic accuracy
both in the Standard Model and in the MSSM.}
\label{fig}
\end{figure}

Starting from the formulae given in the previous Section II, we have
now computed the various electroweak
logarithmic effects on the considered
observables in the two approximate expansions, to subleading
logarithmic accuracy, in an energy region between 500 GeV and 3 TeV,
that should include the foreseen LC and CLIC c.m. energy domains.
To simplify the presentation of our results, we have considered
the known fact that the number of final longitudinal $WW$ pairs is
much smaller than that of the transverse ones, that completely
dominate the available statistics (leaving aside the technical
difficulties of analyzing the final polarization). In this spirit, we
shall present a detailed analysis for the unpolarized final $WW$
state, showing all the considered observables in this case in 
Fig.~(\ref{fig}a,c,d). 
For sake of comparison with previous papers, we also draw 
in Fig.~(\ref{fig}b) the plot
representing the cross section for final longitudinal $WW$ pairs.\par
The results are given both for the SM and for the MSSM.
From a glance to the different
graphs, a number of features emerge. To proceed with order, it is more
convenient to give a separate discussion of the three considered
observables.\\

a) Cross sections. One sees from Fig.~(1a) that the two approximate
expansions for unpolarized $WW$ pairs are
"essentially" ({\em i.e.} within the
assumed 1-2 percent accuracy) identical for c.m. energies below,
roughly, 1 TeV. When the energy increases beyond this value, the
difference between the two approximations becomes larger, reaching a
dramatic (in our opinion) value of ten percent at about 3 TeV. We
observe also quite a small difference between the SM and the 
MSSM effects in
our considered logarithmic expansions, in agreement with the
observation of Ref.\cite{Hahn} for what concerns the one-loop
approximation.\\

As one sees from inspection of Fig.~(1b) identical remarks strictly apply
for the longitudinal $WW$ cross section. We can therefore conclude
that for what concerns cross sections, \underline{at the subleading
electroweak logarithmic accuracy}, a one-loop expansion does not
require extra corrections below one TeV, but appears to be drastically
inadequate in the higher (2,3 TeV) c.m. energy domain.\\

b) Forward-backward asymmetry.  As one immediately sees from Fig.~(1c),
the same conclusions given for the cross sections case are valid. In
particular, at 3 TeV, the difference between the two approximations
reaches a value of five percent, to be compared with a Born value of
approximately 0.77 (with our choice of cut at 30 degrees).
Again, no appreciable difference exists between the SM and the MSSM
results.\\

c) Electron-positron longitudinal polarization asymmetry. This is the
only case where, between the one-loop expansion and the resummed one,
no appreciable difference exists in the full considered energy
range (see Fig.~(1d)).
This is relatively simple to understand since this process is
dominated by left-handed electron contributions, given the
nature of the final state. In fact, not much information seems to be
provided by a measurement of this observable, at least in the
framework of a "conventional" model like the MSSM.

\section{Conclusions }

The main motivation of our work was that of examining the reliability
of a one-loop expansion for the process $e^+e^-\to W^+W^-$ in the MSSM
(and also in the SM). The practical reason is that we do not see
personally many chances of performing a two-loop calculation in this
model (unless a dedicated effort is motivated and supported). Naively,
one reason of worry would be the realization that the one-loop effect
is "large", say beyond the qualitative ten percent threshold, that
might induce the feeling of a two-loop effect possibly
beyond the one-two
percent limit. In this respect we must make a preliminary distinction
between two different types of one-loop effects. The first ones are
the "classical" QED ones. In the case of $WW$ production, they are
known \cite{BD, Hahn} to be large and in some cases dominating the
overall correction. However, they are supposedly under control, and
thus the largeness of their size does not generate particular worries
since an established resummation procedure exists. In addition they
are purely standard, {\em i.e.} the same in SM and in MSSM. A quite different
situation characterizes the "genuine" electroweak effects. 
For the latter
a \underline{complete} resummation procedure of the 
one-loop contribution to all orders
is not known at the moment. Thus, the existence of large one-loop
corrections might lead to the feeling that unknown higher order
effects might be relevant. In the case of $WW$ cross sections, the
Sudakov effects at one loop become indeed, rather quickly, "large". As
one sees, they reach already the fifteen percent level at 1 TeV. The
technical reason of this fact, which is not shared by other (fermion,
scalar) final states, is not difficult to understand from our formulae.
In fact, for final $W$ pairs, the coefficient $b^{fin, TT}$ only
contains a (negative) squared logarithm that is not reduced by a
corresponding linear logarithm like for other final state processes.
This explains the quick rise of the \underline{negative} effect, a
feature that would reappear for final chargino production \cite{NR}.
In particular, at 3 TeV, the size of the one-loop term reaches the
$\simeq 50$ percent value, that seems, honestly, disturbingly
large. To a minor level, an analogous feature characterizes
the considered forward-backward asymmetry.\\

In such a situation, the only (to our knowledge) control is offered by
the existing \underline{partial}
resummation procedures. For the MSSM case, we have used
the only one that we are aware of \cite{revmel} (therefore we cannot
compare it with other proposals). We remind the reader that both the
one-loop Sudakov expansion and the fully resummed one are valid to
\underline{subleading} logarithmic accuracy. Within this limitation, we
have verified that, below 1 TeV, no "dangerous" ({\em i.e.} at the one-two
percent level) difference exists between the two approximations, while
large discrepancies arise at higher energies. We can conclude that,
\underline{to subleading electroweak logarithmic accuracy}, a one-loop
calculation is fully adequate below one TeV, and certainly
unsatisfactory at higher energies. The next remaining step is to
enlarge this statement to include a complete one-loop calculation. In
our opinion, this conclusion is, at least, rather reasonable, given
the fact that the difference between the two approaches (modulo the
known QED corrections) is only due to next-to subleading ({\em e.g.}
constant) terms. From our previous experience in the case of final
charged Higgs production \cite{HH}, we expect that such terms are
reasonably small and sufficient to provide an adequate description of
the process, when added to the logarithmic ones. Certainly, this
statement needs a professional support. In practice, this would be
provided by the comparison of a complete one-loop program with a
logarithmic expansion that includes {\em e.g.} an extra constant term, as we
did thoroughly in Ref.\cite{HH}. This is not beyond a 'reasonable"
effort. We consider this suggestion as a much simpler possibility
compared to the (tough!) alternative of performing a two-loop
calculation, and look forward to its completion in a reasonably near
future.\par
For what concerns the sensitivity of $WW$ production to the genuine
SUSY component of the MSSM, our conclusion is that, at least to
subleading logarithmic accuracy, this appears to be rather small (both
at the one-loop  \underline{and also at the resummed level}) in the
full energy range that we have considered. In fact, at this level, the
only difference between the SM and the MSSM is due to the change of
the coefficients $n_{i,f}$ of the linear logarithms, as explained in
Section 2. Although we cannot exclude a higher sensitivity in the
next-to subleading terms where the parameters of the model will
actually enter, it appears from the previous Hahn analysis \cite{Hahn}
that, at least until 1 TeV, the complete calculation retains this
property. This could indicate, in case of SUSY discovery, that $WW$
production might be more relevant for detecting possible signals of
different types of sophisticated new physics models.

\newpage

\begin{thebibliography}{99}

\bibitem{WWrev} For a recent review see {\em e.g.} K. M\"onig,
hep-ph/0303023 and references therein.

\bibitem{LEP2} The LEP collaborations, CERN-EP/2002-091,
hep-ex/0212036.

\bibitem{LC} see {\em e.g.},
E.~Accomando {\it et.al.}, Phys. Rep. {\bf C299},299(1998).

\bibitem{CLIC}
 " The CLIC study of a multi-TeV $e^+e^-$ linear
collider", CERN-PS-99-005-LP (1999).

\bibitem{BD} W.~Beenakker and A. Denner, Int.J.Mod.Phys.
{\bf A9}(1994)4837.



\bibitem{Hahn} T. Hahn, Nucl.Phys.{\bf B609} (2001) 344 and earlier
references for SM and MSSM cases therein.




\bibitem{sud} V.~V.~Sudakov,
Sov.~Phys.~JETP~3, 65 (1956);
Landau-Lifshits:
Relativistic Quantum Field theory IV tome (MIR, Moscow) 1972;
M. Kuroda, G. Moultaka and D. Schildknecht, Nucl. Phys.
{\bf B350},25(1991);
G.Degrassi and A Sirlin, Phys.Rev. {\bf D46},3104(1992);
W.~Beenakker et al,  Nucl.~Phys.~{\bf B410}, 245 (1993),
Phys.~Lett.~{\bf B317}, 622 (1993);
A.~Denner, S.~Dittmaier and R.~Schuster,
Nucl.~Phys.~{\bf B452}, 80 (1995);
A.~Denner, S.~Dittmaier and T.~Hahn, Phys.~Rev.~{\bf D56}, 117 (1997);
P.~Ciafaloni and D.~Comelli, Phys.~Lett.~{\bf B446}, 278 (1999);
M. Beccaria, P. Ciafaloni, D. Comelli, F. Renard, C. Verzegnassi,
Phys.Rev. {\bf D 61},073005 (2000);
Phys.Rev. {\bf D 61},011301 (2000).



\bibitem{eqth} J.M. Cornwall, D.N. Levin and G. Tiktopoulos,
Phys.Rev.{\bf D10}(1974)1145; G.J. Gounaris, R. K\"ogerler and H.
Neufeld, Phys.Rev.{\bf D34}(1986)3257.




\bibitem{revmel} M.~Melles, Phys. Rep.{\bf 375}(2003)219, and
references therein.

\bibitem{scalar}
M.~Beccaria, M.~Melles, F.~M.~Renard and C.~Verzegnassi,
Phys.Rev.{\bf D65}, 093007 (2002).

\bibitem{HH} M. Beccaria,  F.M. Renard, S. Trimarchi, C. Verzegnassi,
hep-ph/0212167.

\bibitem{susylog}
M. Beccaria, F.M. Renard and C. Verzegnassi,
Phys. Rev. {\bf D63}, 095010 (2001);
Phys.Rev.{\bf D63}, 053013 (2001).

\bibitem{Splitting} see {\em e.g.} G. Altarelli, 
Phys. Rep. {\bf C81}, 1 (1981).
 

\bibitem{real} M.~Beccaria, F.M.~Renard, C.~Verzegnassi,
hep-ph/0203254; Linear Collider note LC-TH-2002-005.


\bibitem{dp1} A.~Denner, S.~Pozzorini; 
Eur.~Phys.~J.~{\bf C18} (2001) 461.




\bibitem{NR} M. Beccaria, M. Melles, F.M. Renard, 
S. Trimarchi, C. Verzegnassi, hep-ph/0304110.



\bibitem{resum}
 M.~Ciafaloni, P.~Ciafaloni, D.~Comelli,
Phys.~Rev.~Lett. {\bf 84}(2000) 4810;
M.~Melles,
Phys.~Lett.~{\bf B495} (2000) 81;
Phys. Rev. {\bf D63} (2001) 034003;
{\bf D64} (2001) 014011; Phys.~Rev.~{\bf D64} (2001) 054003;
Eur.Phys.J.~{\bf C24} (2002) 193;
J.H.~K\"uhn, A.A.~Penin, V.A.~Smirnov; Eur.~Phys.~J.~{\bf C17}
(2000) 97;
J.H.~K\"uhn, S.~Moch, A.A.~Penin, V.A.~Smirnov; Nucl.~Phys.~{\bf
B616} (2001) 286.


\bibitem{twoloop} M.~Hori, H.~Kawamura, J.~Kodaira, 
Phys.~Lett.~{\bf B491} (2000)275;
W.~Beenakker, A.~Werthenbach; Nucl.~Phys.~{\bf B630} (2002) 3.


\bibitem{DMP} V.S.~Fadin, L.N.~Lipatov, A.D.~Martin, M.~Melles;
Phys.Rev. {\bf D61} (2000) 094002; 
A.~Denner, M.~Melles, S.~Pozzorini, PSI-PR-03-01.
TTP-03-03, hep-ph/0301241.



\end{thebibliography}
\end{document}